\documentclass[twocolappendix,  twocolumn]{openjournal}

 \usepackage{amssymb}
 \usepackage{natbib}
\usepackage{graphicx}

\usepackage{xcolor}
 
\usepackage[utf8]{inputenc}
\usepackage[english]{babel}
\usepackage{hyperref}
\hypersetup{
    colorlinks=true,
    linkcolor=blue,
    filecolor=blue,      
    urlcolor=red,
    citecolor=blue,
}
\usepackage{color,colortbl}

 \DeclareGraphicsExtensions{.bmp,.png,.jpg,.pdf}
\usepackage{verbatim}
\usepackage[normalem]{ulem}
\usepackage{orcidlink}
\usepackage{soul}

\begin{document}
   \title{Is the Radcliffe wave  turbulent?}
   \author{Itzhak Goldman $^{1,2}$}
\affiliation{$^{1}$Physics Department, Afeka College, Tel Aviv 6998812, Israel}
\affiliation{$^{2}$ Astrophysics Department, Tel Aviv University, Tel Aviv 6997801 , Israel}

 \begin{abstract}
   Tidal interaction with a satellite galaxy and Kelvin Helmholtz instability were proposed as the generation mechanisms of the Radcliffe wave. In each of these scenarios it is likely that turbulence could be generated.We intend to find out wether there exists a large scale   turbulence, and if so to obtain the turbulence characteristics. We use the observed vertical velocity field, of various young tracers of the gas kinematics, obtained by \citet{ Li+Chen2022},  in order to test for the existence of a large scale turbulence. We do so by computing the autocorrelation function, the  power spectrum and the second order structure function for the two  vertical velocity fields.  The latter suggest   the existence of large scale compressible, Burgers, turbulence.The turbulence timescale on the largest spatial scale is $\sim (400-530)   Myr$, setting  a lower limit on the age of the turbulence and on the age of the Radcliffe wave  . The turbulence region depth   in a direction perpendicular to the   Radcliffe wave direction   is $\sim(500-520) pc.$
\end{abstract}  
\begin{keywords} 
{Galaxy: structure – Galaxy:solar neighborhood-Galaxy:kinematics - Galaxy:turbulence}
  \end{keywords}   
\maketitle

  \section{Introduction}            
    Since its discovery \citep{Alves+2020}, the Radcliffe wave  drew the attention and interest of many researchers. Suggestions regarding its formation mechanism  included tidal interaction with    a satellite galaxy  or star cluster, e. g. \citep{Tepper+2022} and a hydrodynamical instability e.g.  a Kelvin Helmholtz (KH) instability  \citep{Fleck2020}. Following the discovery of the   wave, there were investigations aimed at  obtaining  (in addition to the spatial structure) also kinematic information, e.g.  \citep{Tu+2022, Thulasidharan+2022, Bobylev+2022, Li+Chen2022, Konietzka+2024,  Zhu+2024}. These papers used the vertical velocities   of young   tracers to obtain 
the   velocity  
 of the gas .  The underlying rational was  that the tracers  were created at rest relative to the gas. The young age of the tracers ensures that these velocities have not changed much since they formed.
At   positions along the wave, the vertical velocities of the tracers were derived, and   averaged to obtain a representative value of the vertical velocity at a given position. 

The above papers tried to find out wether the Radcliffe wave is a coherent oscillating structure.  The aim of the present paper is to find out wether the  Radcliffe wave is turbulent. 

While there may exist local drivers of small scale turbulence, we set to find out wether  there exists a large scale turbulence, of a scale comparable to the the size of the Radcliffe wave.  Such a turbulence and its characteristics, could be helpful in understanding the creation mechanism of the Radcliffe wave. 
 
 We employ the observational vertical velocity fields obtained by  \citet{ Li+Chen2022}  This paper gives an average representative observational velocity at each position along the Radcliffe  wave.   The papers of \citet{Zhu+2024} and \citet{Konietzka+2024} provide analytic    fit functions and thus by construction are auto correlated. Therefore, it  cannot be used to test for the existence of auto-correlations of the observational velocities.  The other papers mentioned  above do not provide  a sufficient spatial sampling.
 
 We derive for the two observational vertical velocity fields, the autocorrelation function, the power spectrum and the second order structure function.
 
The paper is organized as follows. In  section 2  we  address the observational vertical velocity fields of   \citet{Li+Chen2022}. In  section 3 we obtain the autocorrelation function of the t  velocity field . In section 4 the power spectrum of the  velocity field is obtained. In section 5 the structure function  of the  velocity field  is derived,  Discussion is presented in section 6. In the appendices  we obtain the structure function and the power spectrum  of a quantity that is the result of integration   in a direction perpendicular to the $x$ axis. The results can be used to estimate the depth of the turbulent region in the perpendicular direction. The computations were done with Wolfram Mathematica 14.3.

\section{The vertical velocity, $v_z$ } 
 We used The Engauge digitizer12.1 to read the values of the position and vertical velocity from the 
 velocity curve of   \citet{Li+Chen2022}.
 
\begin{figure}[h!]
\centering
   \includegraphics[scale=0.5]{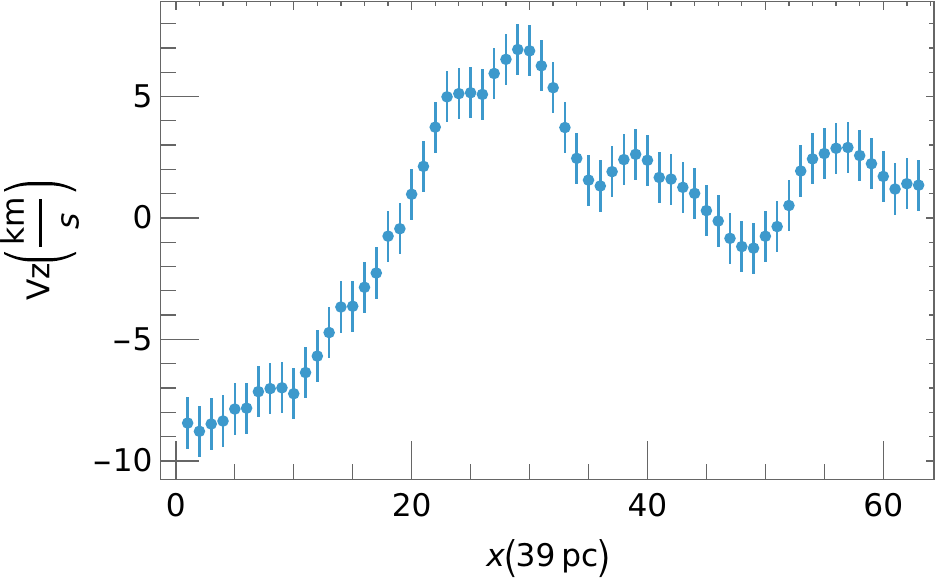}
    \caption {The vertical velocity, of  \citet{Li+Chen2022} , in units   of $km /s$,   as function of position along the wave, in units of $39 pc$. }
  \label{Li+Chen22_vel}
\end{figure}

  These authors used Young Stellar Objects (YSOs) as the vertical velocity tracers of the gas to obtain  the vertical velocity $v_z$ along the direction of the Radcliffe wave. Fig.\ref{Li+Chen22_vel} displays the velocity field after subtraction of its mean value.  The velocity is in units of $km/s$ and the position in units of $39 pc$, The total length along the wave is  $L=2.42$ kpc. The standard deviation of $v_z$ is $4.45 km/s$.
   From the observational plot in \citet{Li+Chen2022}, we estimate an uncertainty  of $ \pm 3 km/s$ in the values of the vertical velocity. Note that this is of the order of the standard deviation of $v_z$.
 
 on of $v_z$ is $5.95 km/s$.

   \section{The autocorrelation function of the vertical velocity field}
   The existence of long range correlations is a basic characteristic of turbulence. In this section we set to obtain the autocorrelation function for the   vertical velocity field  in order to test for the existence of long range correlations. The  autocorrelation is defined as 
  
  \begin{equation}
C(x)=\left <\left (v_z (x') -\bar{v_z}\right )\left (v_z(x'+x) -\bar{v_z}\right ) \right>
\end{equation}
Here, the angular brackets denote averaging over $x'$ and  $\bar{v_z}$ is the mean value of the velocity field.

The uncertainties of the autocorrelation function were evaluated by generating 300 velocity fields in which the velocity at a given position is    randomly displaced from the observational velocity and is within the observational uncertainty. The autocorrelation function of each of the simulated velocity fields is computed.
  Then, the standard deviation of the resulting ensemble of the autocorrelation functions is  taken as the sought for uncertainty. 
 
\begin{figure}[h!]
\centering
   \includegraphics[scale=0.5]{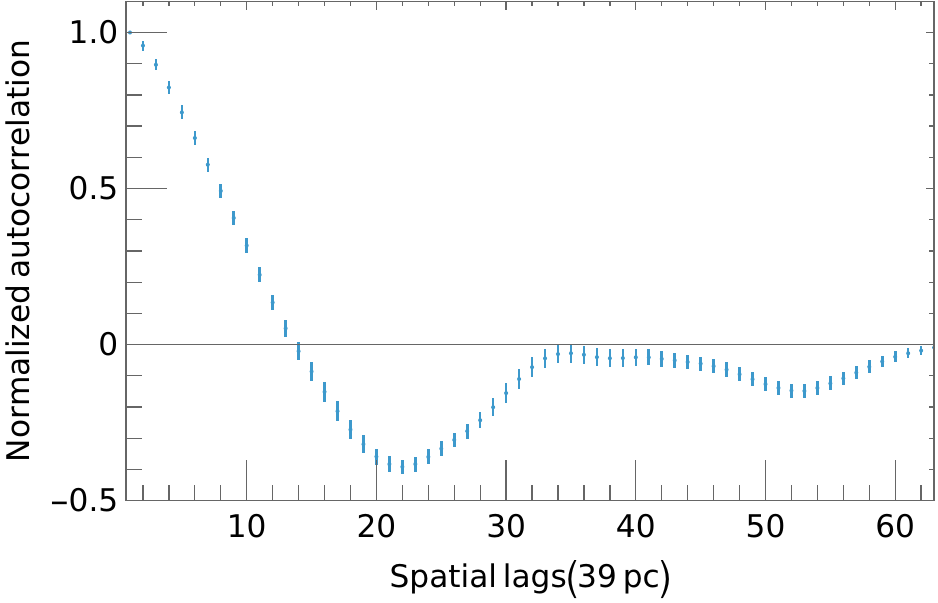}
    \caption {The normalized autocorrelation function of the vertical velocity, of  \citet{Li+Chen2022} ,  as function of position along the wave, in units of $39 pc$. }
  \label{Li+Chen22_ac2}
\end{figure}

 Fig.\ref{Li+Chen22_ac2}  displays the normalized autocorrelation function  of the    velocity field  of \citet{Li+Chen2022}.  . It is evident that the velocity field  possesses  correlations extending over the entire spatial length of the data. The computed error bars  turn out to be very small,
 
\section{The power spectrum of   the vertical velocity field} 
 
 In this section we evaluate the discrete power spectrum for the   observational velocity field . It is equal to the squared absolute value of the  discrete Fourier transform of the vertical velocity field.
 
 The uncertainties of the power spectrum  were evaluated in the following procedure. We generated  300 simulated  vertical velocity fields in which the velocity at a given position is    randomly displaced from the observational velocity and is within the observational uncertainty. The power spectrum of    each of the simulated vertical velocity fields is computed and the   standard deviation of the resulting ensemble of the power spectra is  taken as the  uncertainty of the power spectrum.
 
 In the   following, the power spectra are plotted as function of the relative, dimensionless, wavenumber $q=k/k_0= k L/(2\pi)$ 
where k is the wavenumber,  $k_0$ is the smallest wavenumber, and $L$ is the the largest spatial scale.

\begin{figure}[h!]
\centering
   \includegraphics[scale=0.5]{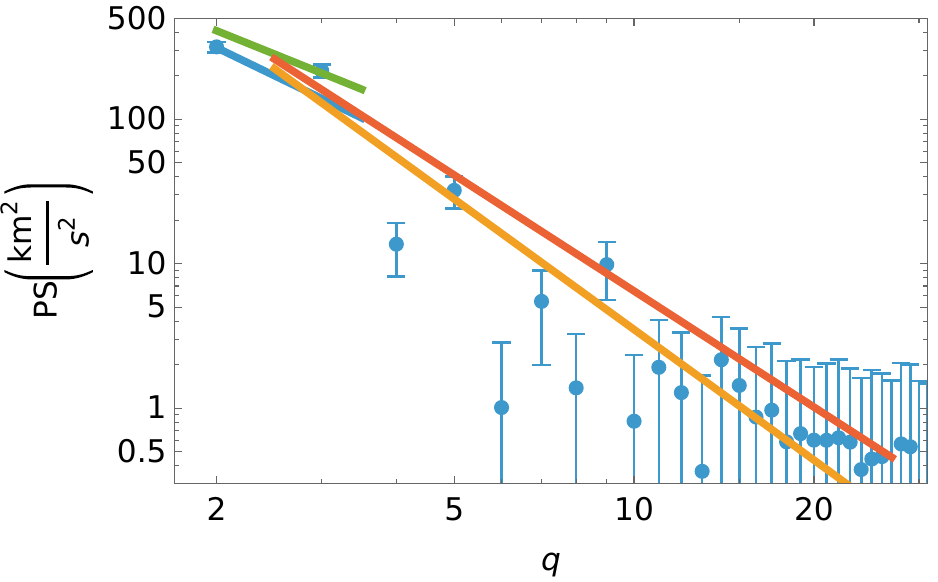}
    \caption {The 1D Power spectrum  of $v_z(x)$ of Fig. \ref{Li+Chen22_vel},  in units   of $km^2/s^2$    as function of  the dimensionless wavenumber q. The blue line has a logarithmic slope of -2, the orange line has a logarithmic slope of -3, the green line has a logarithmic slope of -5/3, and the red line has a logarithmic slope of -8/3.} 
  \label{Li+Chen22_ps3}
\end{figure}

Fig. \ref{Li+Chen22_ps3} displays the power  spectrum of the vertical velocity field  of  \citet{Li+Chen2022}  
The blue line has a logarithmic slope of -2, the orange line has a logarithmic slope of -3, the green line has a logarithmic slope of -5/3 and the red line has a logarithmic slope of -8/3. 

 The blue and orange lines are asymptotes corresponding to  compressible turbulence while the green and red lines  are asymptotes corresponding to Kolmogorov turbulence. Inspection of the figures suggests that compressible turbulence describes better the power spectra,
 however, the difference is not large. 

  \section{The structure function of the vertical velocity field} 
 
 The second order structure function is defined as
 
 \begin{equation}
 S(x)= <\left( v_z(x+x') - v_z(x')\right)^2>
 \end{equation}
  
 where the angular brackets denote averaging over $x'$. 
 The uncertainties of the structure function  were evaluated in the following procedure. We generated  300 simulated  vertical velocity field  in which the velocity at a given position is    randomly displaced from the observational velocity and is within the observational uncertainty. The structure function  of each of the simulated vertical velocity fields is computed and the   standard deviation of the resulting ensemble of the structure functions is  taken as the  uncertainty of the structure function. 
  
\begin{figure}[h!]
\centering
   \includegraphics[scale=0.5]{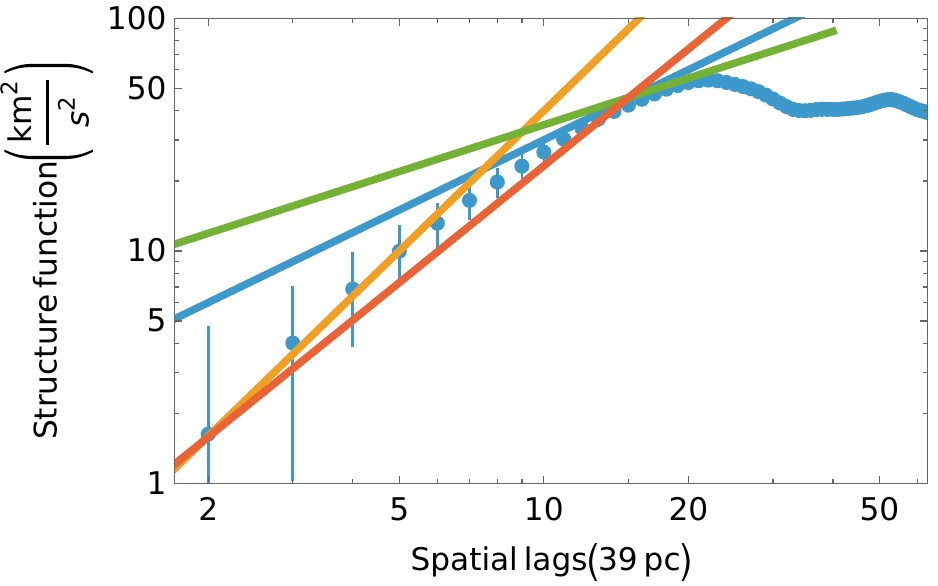}
    \caption {The structure function of  $v_z$ of Fig. \ref{Li+Chen22_vel}, in units   of $km^2/s^2$    as function of  spatial lags in units of $39 pc$. The blue line has a logarithmic slope of 1, the orange line has a logarithmic slope of 2, the green line has a logarithmic slope of 2/3, and the red line has a logarithmic slope of 5/3.}
       \label{Li+Chen22_sf3}
\end{figure}

 Fig.\ref{Li+Chen22_sf3}  displays the structure function  for the   vertical velocity field of  \citet{Li+Chen2022} . In   The blue line has a logarithmic slope of 1, the orange line has a logarithmic slope of 2, the green line has a logarithmic slope of 2/3, and the red line has a logarithmic slope of 5/3.

 The blue and orange lines are asymptotes corresponding to  compressible turbulence while the green and red lines  are asymptotes corresponding to Kolmogorov turbulence. Inspection of the figures suggests that compressible turbulence describes much better the structure functions,
 
\section{Discussion}

The goal of the present paper was to test wether a large scale turbulence exists in the Radcliffe wave. There may be local drivers of small scale turbulence but these are not the subject of the present paper. 
 
The autocorrelation function of each of the observational vertical velocity fields exhibits long range correlations extending over the entire spatial extent. This is consistent with turbulent velocity fields.   The normalized autocorrelation functions are quite similar,
 
  Asymptotes according to compressible, Burgers, turbulence and incompressible , Kolmogorov, were added to the figures of the power spectrum and structure function,   

The power spectrum suggest compressible turbulence but  Kolmogorov turbulence cannot be ruled out. However, the structure function s
are not consistent with Kolmogorov turbulence but are consistent with compressible turbulence.

   Compressible turbulence power spectra were observed in HI  intensity maps in the Milky Way (MW) galaxy \citep{Green93} and in the SMC \citep{Stanimirovic+99}. This power spectrum has been observed also  in molecular clouds \citep{Larson81,Leung+82,Dame+86}, and in the   HII region Sharpless~142  \citep{Roy+Joncas85}. It has been found in a shocked cloud near the Milky Way galaxy center \citep{Contini+Goldman2011},   in the Gamma ray emission from the large Magellanic Cloud \citep{Besserglik+Goldman2021}, and recently in CSWA13, a  gravitationally lensed Star-Forming galaxy at $z = 1.87$ with outflowing wind \citep{Goldman2024}. It has been   obtained also in numerical simulations e.g.  \citep{Passot+88,Vazquez-Semadeni+97, Kritsuk+2007,Federrath+2021}.

 It has been noted by   \citep{Stutzki+98, Goldman2000,  Lazarian+Pogosyan2000,   Miville+2003a} that when the lateral spatial scale is smaller than the depth of the turbulent layer (in a direction perpendicular to the 1D axis of the turbulence), the turbulence changes from a 1D to a 2D one. Consequently,  the logarithmic slope of the power spectrum 	 steepens   by $-1$ compared to its value when the lateral scale is large compared to the depth. The logarithmic slope of the structure function increases by 1. This behavior has been indeed revealed  in  observational power spectra  of Galactic and extra Galactic turbulence ( e.g. 
\citet{elmegreen+2001} ,\citet{ Block+2010}, \citet {Miville+2003b}, \citet{Goldman2021} ) and in solar photospheric turbulence \citep{Abramenko+Yurchyshyn2020}.

This behavior is also exhibited in the power spectra and the structure functions derived here.
  
In the appendices we obtained  estimates for the depth of the turbulence in the case when  the observed  quantity (the vertical velocity in the present case) is a result of integration in a direction perpendicular to the 1D axis of the turbulence.This would be relevant when the velocity field is determined by an emission or absorption line, In the present case the   velocities represent tracers at different perpendicular distances from the wave direction - but are not strictly an integration in the perpendicular direction.
Nevertheless, it maybe of interest to apply the theoretical relations to obtain an estimate for the depth of the turbulence.

For the case of compressible turbulence. 
the structure function yields $D= 1.83 x_{tr}$, where
 $x_{tr}$ is the value of the spatial lag of the structure function which marks the transition between   logarithmic slopes of $1$ and $2$. 
 
The power spectrum  yields $D =0.49 \frac{L}{q_{tr}}$. Here, L is the largest spatial scale and $q_{tr}$ is the value of $q$, at which the logarithmic slope changes from -2 to -3.

The two power spectra show  a transition at $q_{tr}\sim 2.4$ implying $D\sim 500 pc$, The structure functions change the logarithmic stope at a spatial lag $ x_{tr} \sim 280 pc$ yielding $D \sim 520 pc$. The estimated depths are consistent.

The turbulence timescale on the largest spatial scale is $\tau_{turb}\sim L/v_{z,turb} $., where $v_{z,turb}$ is the r.m.s. turbulent velocity.  For the data of \citet{Li+Chen2022} it is  $4.45 km/s$ resulting in $\tau_{turb}\sim 530 Myr$.  The r.m.s. turbulent velocity for the data of \citet{Konietzka+2024} is  $5.95 km/s$ resulting in $\tau_{turb}\sim 400 Myr$.

  The buildup time  of the turbulence cannot be smaller than the turbulence timescale of the largest spatial scale. Therefore, this estimate represents  a lower limit on  the look- back   time to the time  when the turbulence was generated and hence on the age of the  Radcliffe wave itself.  This information could help decide between possible generation mechanisms of the Radcliffe wave. 
  
  The age of the turbulence is indeed much larger than the age of the tracers used to obtain the velocity field, justifying the assumption that the obtained velocities reflect the velocities of the gas.

\begin{acknowledgements}
 I thank the rersearch aautyhority at Afeka College
\end{acknowledgements} 

 \bibliographystyle{aasjournal}    
 \bibliography{RW_turb}

\appendix 

     \section{The theoretical 1D structure function of a quantity that is the result of integration along a perpendicular direction}

       Consider a function $f(x)$ (where $x$ is a straight line)   which is  an integral  along a perpendicular direction to that line, y:  
      
     \begin{equation}
      f(x) = \int_0^D g(x, y) dy.
      \end{equation}
      Here, $y$ is  the perpendicular coordinate and  $D$ the depth. A plane parallel geometry is assumed for simplicity.

  The autocorrelation function of $f(x)$ (with zero mean value) is: 
  
  \begin{eqnarray}  
  C_f (x)= <f(x+x') f(x)> =\\
   \int_0^D\int_0^D<g(x', y) g(x+x', y') > dy dy'=\nonumber\\
   \int_0^D\int_0^D C_g(x, y-y')dy dy'.\nonumber
    \end{eqnarray} ,
  
 The autocorrelation function  $C_g(x, y-y')$ can be expressed by the two-dimensional power spectrum, $P_2(k_x, k_y)$,
 \begin{eqnarray}
  C_g(x, y-y')= \int_{-\infty}^{\infty} \int _{-\infty}^{\infty}e^{i( k_x x +k_y ( y-y'))   } 
    P_2(k_x, k_y) dk_x  dk_y.   \nonumber 
    \end{eqnarray}
  leading to
   \begin{equation}
    C_f (x, D)=\int_{-\infty}^{\infty} \int _{-\infty}^{\infty}e^{i k_x x}    P_2(k_x, k_y) \frac{  \sin^2  \left( k_y D/2  \right)}       {\left(   k_y D/2  \right)^2 } dk_x dk_y.
    \end{equation} 

From the definition of the   structure function of $f(x)$
\begin{equation}
S_f(x, D)= <\big (f(x+x') - f(x')\big)^2 > \nonumber
\end{equation}
 and Eq. (A.3) follows

\begin{eqnarray}  
\hskip -0.5cm S_f(x, D) 
 \propto  \int_0^{\infty} \int_0^{\infty} \sin^2(k_x x/2)  \frac{\sin^2  \left( k_y D/2  \right)}{  \left(   k_y D/2  \right)^2}\\ 
   P_2(k_x, k_y) dk_x dk_y. \nonumber
\end{eqnarray} 

 In the case of a turbulence with a one-dimensional power spectrum  which is a power law with index $-m$, the two dimensional power spectrum is
 \begin{equation}
 P_2(k_x, k_y) \propto   \left(k_x^2 +   k_y^2\right)^{-(m+1)/2 .}\ \ \  \ \ \  
 \end{equation}
 resulting in
 
   \begin{eqnarray}  
 S_f(x, D) \propto  \int_0^{\infty} \int_0^{\infty} \sin^2(k_x x/2)  \frac{\sin^2  \left( k_y D/2  \right)}{  \left(   k_y D/2  \right)^2}\\
 \nonumber
  \left(k_x^2 +   k_y^2\right)^{-(m+1)/2 } dk_x dk_y. 
\end{eqnarray} 

It is convenient  to define the dimensionless variables
\begin{equation}
 \eta= k_y D/2. \ \ \ \ \  \ \ ; \ \ \ \mu= k_x D/2.  \nonumber
\end{equation}

     \begin{figure}[ht!]
    \centering 

   \includegraphics[scale=0.4] {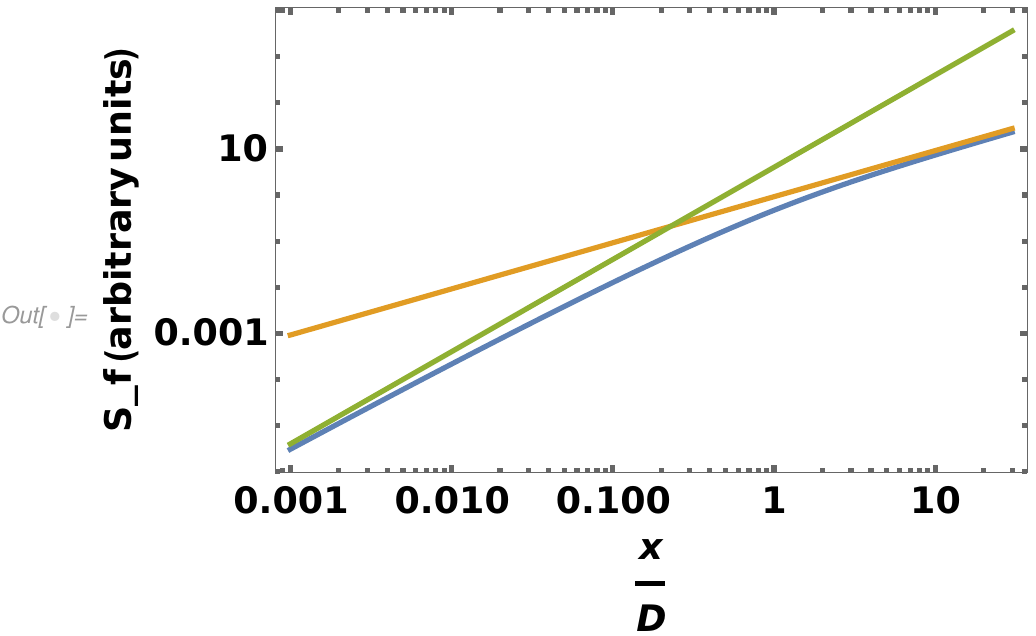 }
   \caption{Theoretical 
    structure function  of  a quantity, which is an integral  
      along  a perpendicular direction to $x$, as function of $x/D$. The   orange line has logarithmic slopes equaling 1 and the green line has a logarithmic slope of 2.}  
     \label{sf_th1}
     \end{figure} 
     
      \begin{figure}[ht!]
    \centering
   \includegraphics[scale=0.4] {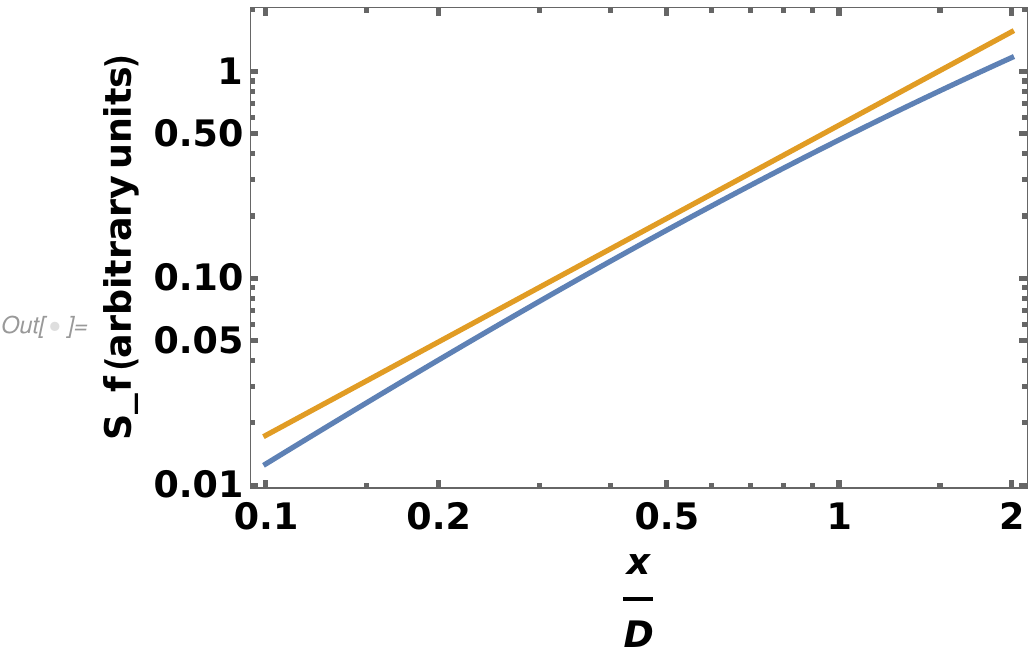 }
   \caption{Theoretical  
    structure function   
    of  a quantity, which is an integral along  a perpendicular direction to $x$, as function of $x/D$. The straight line has a logarithmic slope of 1.5.}  
     \label{sf_th2}
     \end{figure}   
     
     The structure function of equation (A.6) can be expressed as       
         \begin{equation}                        
  S_f(x ,D)
   \propto \int_0^{\infty}I(\mu)\sin^2\left( \mu x/D  \right)    d\mu.
      \end{equation}
  
 where  $I (\mu)$ is
 \begin{equation}
        I(\mu )=   \int_0^{\infty}  \left(\mu^2  +  \eta^2\right)^{-(m+1)/2} \frac{ \sin^2 \eta}{     \eta^2} d\eta.
   \end{equation}

Eq. (A.7)    implies  that the structure function  argument is $x/D$. 
Also, inspection of   Eq. (A.7) and eq.  (A.8) reveals that for $x<<D$ the structure function is proportional to $x^m$ while for $x>>D$
it is proportional to  $x^{m -1}$ .
 
 A numerical solution for the case of $m=2$ is presented in Fig.\ref{sf_th1} together with  power laws with exponents 1 and 2. 
  In order to find the value of $x_{tr}/D$, where  $x_{tr}$ denotes the transition lag between the two slopes,   a power law with exponent of 1.5 is plotted in  Fig.\ref{sf_th2} together with the structure function. The value of $x_{tr}/D$ is taken to be   the value for which the logarithmic
 slope of the structure function equals 1.5, i.e. where the straight line  is tangent to the structure function.
  The result is $x_{tr}/D \sim 0.547$; thus   $D \sim 1.83 x_{tr}$.
 
   \section{ The theoretical power spectrum of a quantity that is the result of integration along a perpendicular direction} 
 \label{sec:app_The_theoretical_power_spectrum}

The 1D power spectrum $P_f(k_x)$ is by definition

\begin{equation}
P_f(k_x)= \int_{-\infty}^{\infty} e^{-ik_x} C_f(x) dx
\end{equation}
\\
From Eq. (A.3) and Eq. (A.5 follows

\begin{equation}
P_f(k_x) \propto \int_{-\infty}^{\infty}\left(k_x^2 +   k_z^2\right)^{-(m+1)/2 }  \frac{\sin^2  \left( k_z D/2  \right)}{  \left(   k_z D/2  \right)^2} dk_z 
\end{equation} 
\\
Using the definitions of $\eta$ and $\mu$ as well as  Eq. (A. 8)  we obtain
\begin{equation}
P_f(k_x) = P_f(\mu)= I(\mu)
\end{equation}
 
\begin{figure}[ht!]
 \centering
   \includegraphics[scale=0.4] {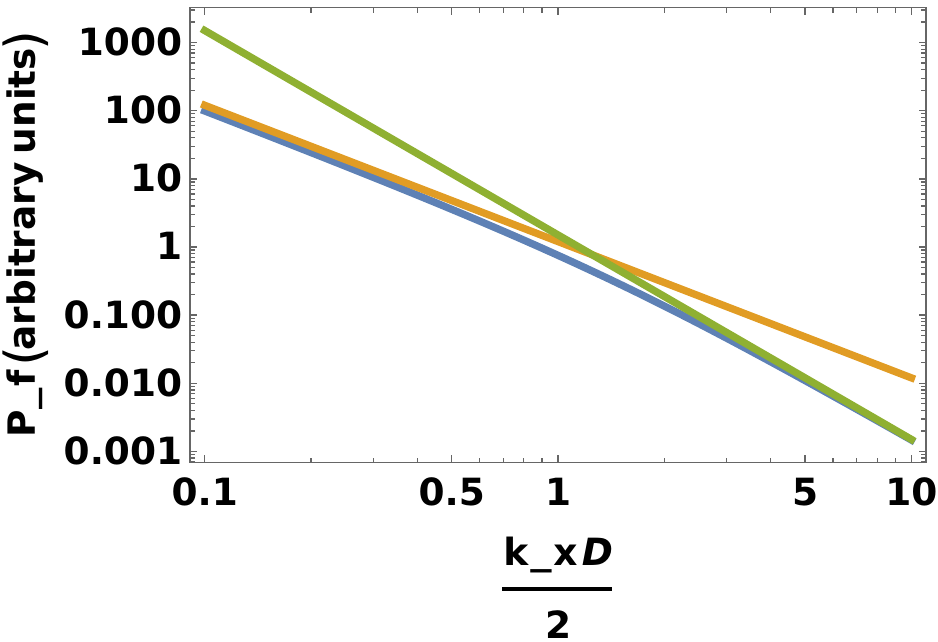 }
   \caption{Theoretical 
power spectrum, of a quantity which is an integral  
      along a perpendicular direction to $x$, as function of $\mu =k_x D/2$. The orange line has a logarithmic slopes equaling -2 and the green line has logarithmic slope of -3.}  
 \label{ps_th1}
     \end{figure}

Thus, the power spectrum is a function of $\mu$. Inspection of Eq. (A.8) and Eq. (B.3) reveals that for 
$\mu<<1$, $P_f( k_x)\propto k_x^{-m}$, while for $ \mu>>1$,  $ P_f( k_x)\propto k_x^{-m-1} $. For the present case of $m=2$ the power spectrum is plotted in Fig.\ref{ps_th1}. Plotted also are the asymptotes 
with logarithmic slopes of -2 and -3.

\begin{figure}[ht!]
 \centering
   \includegraphics[scale=0.4] {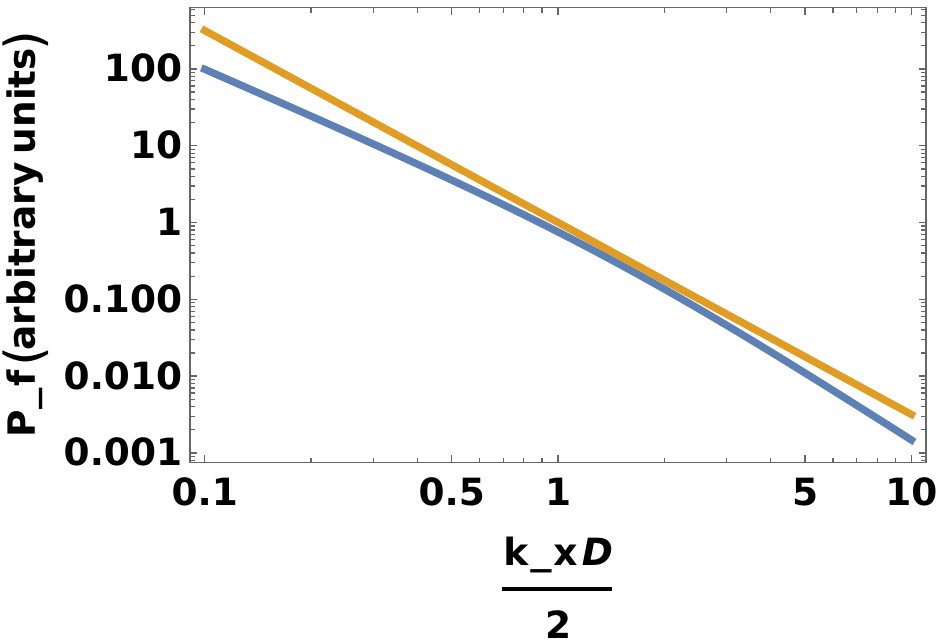 }
   \caption{Theoretical 
power spectrum, of a quantity which is an integral  along a perpendicular direction to $x$, as function of $\mu=k_x D/2$. The straight line has a logarithmic slope equaling -2.5.}  
 \label{ps_th2}
     \end{figure}

In order to find the transition value $\mu_{tr}$ a line with a logarithmic slope of -2.5 is plotted and the value of $\mu_{tr}$ is taken to be the tangent point to the power spectrum. From Fig.\ref{ps_th2}
we find that $\mu_{tr}=\frac{D}{2} k_{x,tr}  =1.54$. implying
\begin{equation}
D= 0.49\frac{L}{q_{tr}}
\end{equation}
Here, $q_{tr}= k_{x, tr}(\frac{L}{2\pi})$ is the  value of the relative dimensionless wavenumber that marks the transition between a logarithmic slope of -2 and a logarithmic slope of -3.

\end{document}